# Carbon Kagome lattice and orbital frustration-induced metal-insulator transition for optoelectronics

Yuanping Chen[1,2], Y.Y. Sun[1], H. Wang[1], D. West[1], Yuee Xie[2], J. Zhong[2], V. Meunier[1], Marvin L. Cohen[3] and S. B. Zhang[1*]

[1]Department of Physics, Applied Physics, and Astronomy Rensselaer Polytechnic Institute, Troy, New York, 12180, USA.
[2]Department of Physics, Xiangtan University, Xiangtan, Hunan 411105, People's Republic of China.
[3]Department of Physics, University of California at Berkeley, and Materials Sciences Division, Lawrence Berkeley National Laboratory, Berkeley, California, 94720, USA.

A three-dimensional elemental carbon Kagome lattice (CKL), made of only fourfold coordinated carbon atoms, is proposed based on first-principles calculations. Despite the existence of 60° bond angles in the triangle rings, widely perceived to be energetically unfavorable, the CKL is found to display exceptional stability comparable to that of $C_{60}$. The system allows us to study the effects of triangular frustration on the electronic properties of realistic solids, and it demonstrates a metal-insulator transition from that of graphene to a direct gap semiconductor in the visible blue region. By minimizing *s-p* orbital hybridization, which is an intrinsic property of carbon, not only the band edge states become nearly purely frustrated *p* states, but also the band structure is qualitatively different from any known bulk elemental semiconductors. For example, the optical properties are similar to those of direct-gap semiconductors GaN and ZnO, whereas the effective masses are comparable or smaller than those of Si.

Kagome lattices, a triangular lattice of spacing *a* after eliminating every fourth site from its triangular sub-lattice of spacing *a*/2, have attracted attention for studying the physics of frustration [1-5]. They were first studied because of the spin frustration properties of anti-ferromagnetic spin states [6], where a spin cannot find an orientation which simultaneously favors all the spin-spin interactions with its neighbors. Spin frustration is known to yield fascinating effects such as the formation of spin ices, spin liquids, and spin glasses [1,2,6,7]. Frustration can also lead to new fundamental phenomena such as the fractional quantum numbers [8], magnetic monopoles [9] and exotic forms of superconductivity [10].

Beyond spin, other fundamental physical quantities such as electronic orbitals should also show frustration in a Kagome lattice. For example, a *p*-orbital can be viewed as a rank one tensor (i.e., a vector) with a clearly defined polarity pointing from its negative lobe to positive lobe [11-15], in analogy to a spin vector. Figure 1 shows a schematic diagram of the *p*-orbital frustration in a two-dimensional (2D) Kagome lattice made of three lattice sites (I, II, and III) in a primitive cell. Here, each lattice site hosts three orthogonal *p*-orbitals -- two in-plane, namely, $p_1$ in Fig. 1(a) and $p_2$ in Fig. 1(b), and one out-of-plane, namely, $p_3$ in Fig. 1(c). A linear combination of the *p* orbitals, $p_1$, $p_2$, or $p_3$, at the different lattice sites forms the states for the lattice. However, no matter how one chooses the polarities at the lattice sites, orbital frustration will always occur.

The difference between spin frustration and orbital frustration is that the spin-spin interaction (although weak) can affect ordering of magnetic atoms several lattice distance away, while the orbital-orbital interaction is predominantly between nearest neighbor atoms. Experimentally available Kagome lattices to date are either organic complexes [16-20] or optical lattices [4,21,22]. In these systems, spin frustration and its related physical phenomena have been studied and observed extensively. However, orbital frustration has never been reported because the direct orbital interactions are hindered. As such, the study of orbital frustration is still at the model level [23], despite the fact that it can reveal a potentially different kind of frustration physics than that of the spin case.

In this Letter, a three-dimensional (3D) elemental carbon Kagome lattice (CKL) is proposed, whose stability is comparable to the fullerene $C_{60}$ [24]. This investigation is performed at the crossroad between the physics of orbital frustration with that of graphene and graphene-derivatives [25-28]. We show that orbital frustration in this system not only exists, but it is also responsible for the emergence of a direct band gap of 3.43eV at the $\Gamma$ point of the Brillouin zone (BZ), as determined by hybrid-functional calculations. Our study of the transition of a zero-gap interpenetrated graphene network (IGN) for which no orbital frustration takes place to a fully frustrated CKL reveals that the gap is a direct result of frustration-induced metal-insulator transition. We further show that the electronic and optoelectronic properties of the CKL are excellent for applications since both the electron and hole effective masses are comparable to those of Si, while the imaginary part of the dielectric function, which determines the optical properties, is similar to that of direct-gap GaN and ZnO.

The atomic structure of the CKL is shown in Fig. 2(a). Its hexagonal lattice belongs to space group $P6_3/mmc$ ($D_{6h\text{-}4}$). Each unit cell contains six C atoms, which form two separate equilateral triangles symmetrically placed with respect to an inversion center at the center of the unit cell. The relationship between the 3D CKL and the 2D Kagome lattice is shown in Fig. 2(b) -- if one envisions that each infinitely long zigzag chain in the 3D structure is collapsed to a lattice point, then the two structures become identical.

Our first-principles calculations were based on density functional theory with the PBE approximation [29] to the exchange-correlation functional. The core-valence interactions were described by the projector augmented-wave (PAW) potentials [30] as implemented in the VASP code [31]. Plane waves with a kinetic energy cutoff of 550 eV were used as the basis set. The calculations were carried out in periodic supercells. All atoms were relaxed until the forces were smaller than 0.01 eV/Å. A $9\times9\times16$ k-point set was used for the BZ integration. The band gap was also calculated by using the HSE06 approximation [32]. The frequency-dependent dielectric matrix was calculated following the method described by Gajdos et al. using PAW potentials [33].

In the optimized unit cell of the CKL in Fig. 2(a), the lattice constants are $a$ = 4.46 Å and $c$ = 2.53 Å, respectively. Each C atom forms four bonds. The intra-triangle bond length is 1.53 Å, while the inter-triangle bond length is 1.50 Å. Both values are between the calculated bond lengths of diamond (1.55 Å) and graphene (1.43 Å). Among the six bond angles formed on each C, one (within the triangle) is precisely 60°, one is 115°, and four are 118°. Except for the 60° case, most of the angles are between those of graphene and diamond.

The CKL is made of triangularly-shaped building blocks. Intuitively, one would anticipate it to be unstable because of the large strain [34,35]. However, the calculated cohesive energy $E_{\text{coh}}$ = 7.44 eV/C is large and comparable to that of a $C_{60}$ molecule (7.48 eV/C), although both are lower than graphite (7.85 eV/C) and diamond (7.72 eV/C) [36]. The large $E_{\text{coh}}$ and the calculated large bulk modulus of 322 GPa indicates that the CKL should be a metastable carbon allotrope. The stability originates from the fact that the CKL is a strongly interlocked network of graphene sheets in all three directions, G-1, G-2, and G-3 in Fig. 2(b). Alternatively, the CKL can also be obtained by laterally compressing another more stable carbon allotrope, termed interpenetrated graphene network (IGN) in Fig. 2(c) ($E_{coh}$ = 7.62 eV/C) [36], in the direction of the arrows. The IGN is topologically identical to the CKL, in the sense that the IGN is also made of three zigzag chains. The main difference is that one pair of the C-C bonds in the IGN (see the black dotted lines in Fig. 2(d)) is broken. However, after compression the broken bonds join to form a CKL. To further check the stability of the CKL, we have calculated the phonon spectrum as shown in Supplemental Material Fig. 1 [37], and we did not find any soft phonon mode over the entire BZ.

Figure 3(a) shows the band structure for the CKL. In contrast to all known carbon allotropes and other elemental semiconductors, the CKL has a direct band gap at the Γ-point of the BZ. The PBE gap is 2.35 eV, whereas the HSE gap is 3.43 eV, which is more accurate in most instances [38, 39]. The band structure further suggests large dispersions near Γ for holes and electrons and hence reasonably low masses and large intrinsic mobility. Figure 3(b) shows that the CKL has effective masses comparable to today's electronic flagship material, Si, and a lower hole effective mass than both GaN and ZnO. In particular, the small in-plane effective masses (i.e., $m_e^{//}$, $m_{hh}^{//}$ and $m_{lh}^{//}$) make the CKL a potential wide-gap material for planar electronic devices such as power field-effect transistors (FETs). Figure 3(c) shows the calculated optical properties for CKL, namely, the imaginary part of dielectric function $\varepsilon_2(E)$. Optical transitions between the band edge states are dipole allowed, making the CKL a truly direct gap material similar to GaAs. Furthermore, the absorption is comparable to GaN and ZnO, indicating that CKL could also be well suited for blue to ultra-violet optoelectronic applications.

As the electronic structure of CKL is unique among known elemental semiconductors, it is highly desirable to understand its physical origin, which, as detailed below, is rooted in the *p*-orbital frustration of the Kagome lattice. First, we note from the calculated partial density of states (PDOS) in Fig. 3(d) that the states near the CKL band edges have predominantly carbon *p* character. In contrast, the *s* orbitals are either deeply buried inside the valence band or relatively high inside the conduction band. Being spherically symmetric, *s* orbitals are also not much affected by the lattice frustration. Therefore, for near-band-edge states we may not have to consider the *s* orbitals. Instead, we construct the wavefunctions with appropriate symmetry states out of the *p* orbitals in Fig. 1. The rationale is that by constructing states that obey the character table for the symmetry (see Supplemental Material Table 1 [37]), we might be able to compare the results directly with those calculated for the CKL.

All the wavefunctions corresponding to Table 1 given in the Supplemental Material have been generated. Here, as an illustration, we show how to construct them from the $p_2$ orbitals in Fig. 1, which have been denoted as $\varphi_1$, $\varphi_2$ and $\varphi_3$ on atoms I, II, and III, respectively. By applying the $D_{6h}$ point-group symmetry operations one by one to the orbitals, we find that $p_2$ belongs to either the $B_{1u}$ or $E_{1u}$ irreducible representations of the point group. The corresponding configurations are $\psi_1 = \varphi_1 + \varphi_2 + \varphi_3$, $\psi_2 = 2\varphi_3 - \varphi_1 - \varphi_2$, and $\psi_3 = \varphi_1 - \varphi_2$, as shown in Figs. 1(d-f) (see details in the Supplemental Material [37]). The $\psi_1$ wavefunction is a singlet with antibonding character. This means that, on a cyclic path connecting I, II, and III in Fig. 1(d), $p_2$ always has its head (yellow) follow its tail (blue). The $\psi_2$ and $\psi_3$ wavefunctions are doublets with both bonding and antibonding characters. Hence, the energy of $\psi_1$ is higher than that of $\psi_2$ and $\psi_3$. The wavefunctions for the $p_1$ and $p_3$ orbitals have been constructed in a similar manner. They all have one singlet and one doublet in a group of three, and all the states are orbitally frustrated. However, here the energy of the doublet is higher than that of the singlet because the latter is only composed of bonding states.

As mentioned earlier, atoms in a 3D CKL are not exactly those of a 2D Kagome lattice. Rather, an infinitely long zigzag carbon chain in the former corresponds to a single point in the latter. The band structure of the zigzag chain, with two-fold coordinated carbon atoms, is shown in Fig. 4(a). It can be easily related back to the energy levels of a single carbon atom as the energy dispersion here is solely a result of the C-C interaction in the chain direction to split the levels of non-interacting atoms into corresponding bands. It is interesting to note that the chain is a Dirac metal similar to graphene but with the Dirac point at an arbitrary low-symmetry position.

The important states in Fig. 4(a) are the $\Gamma_1$ and $\Gamma_2$ states, which progressively evolve into the band edges states of the CKL, as can be seen in Figs. 4(b) to 4(f). The corresponding wavefunctions are given at the left panels of Figs. 4(g) and 4(h). With reference to Fig. 1, $\Gamma_1$ and $\Gamma_2$ correspond to the $p_2$ orbitals. We focus next on the CKL band structure in Fig. 4(f) where the band edge states are labeled along with other relevant states. The wavefunctions for these states are given in the right panels of Figs. 4(g) and 4(h). One can see correspondence in the characteristic features of the wavefunctions between $\Gamma_1$ and $\Gamma_{11}$, $\Gamma_{12}/\Gamma_{13}$, and between $\Gamma_2$ and $\Gamma_{21}$, $\Gamma_{22}/\Gamma_{23}$, in the same way the orbital $\varphi$'s correspond to the orbital configuration $\psi$'s in Fig. 1. Note that the zigzag chain is a one-dimensional metal, whereas the CKL is a 3D insulator (or wide gap semiconductor). The zigzag chain is orbitally not frustrated, whereas the CKL is. These mapping suggests an orbital frustration induced metal-insulator transition as elaborated further below.

We recall that the IGN is structurally related to CKL in the sense that they are both made of three zigzag chains. The IGN is not frustrated and its band structure can be easily understood as a result of splitting the bands of the isolated zigzag chains, as can be seen going from Fig. 4(a) to Fig. 4(b). For the IGN, there is no frustration related degeneracy at the $\Gamma$ point. Interestingly, the IGN is also a Dirac metal with low-symmetry and unrelated Dirac points at different places of the BZ (only one of which is shown in Fig. 4(b)). Figures 4(b) to 4(f) show that by applying stress as indicated in Fig. 2(c), the Dirac point is pushed away from the region near the $A$ point of the BZ towards $\Gamma$. It reaches the $\Gamma$ point before the system reaches full frustration, at which point the gap opens. The movements of the corresponding states in Fig. 4 show clearly, when the gap must open, in order to fulfill the degeneracy requirement of the Kagome lattice. Note that our analytical scheme here starting with the zigzag carbon chain is universal; it can easily explain the Dirac metal nature of the graphene lattice. It may also be applied to diamond but a detailed analysis is outside the scope of this study.

Although experimental realization of the CKL has not yet been achieved, our foregoing analysis demonstrates its kinetic stability. The kinetic stability may be sufficient for carbon systems, as a rich number of metastable carbon allotropes have been experimentally fabricated including graphene, nanotubes, and fullerenes. To fabricate the CKL, in particular, one may consider two possible routes (see Supplemental Material Fig. 2 [37]): (1) the elemental building unit of triangular carbon, in the form of different cyclopropane molecules, exists [41,42]. Thus, one may tailor the ligand chemistry of the cyclopropane to realize self-assembly of 2D CKL, similar to the recent success in fabricating metastable carbon nanowiggles [43,44], and then stack them. Our calculation shows that the 2D CKL is a single-layer structure. (2) One may also first consider self-assembly of the IGN, as it has much lower formation energy than the CKL. One can then stress the IGN into a CKL, as discussed earlier. The calculated stress for the transition is about 3 GPa, which should be easily achieved experimentally.

In summary, first-principles calculations predict the existence of triangular-lattice elemental solids. In the case of CKL, remarkable stability, comparable to that of $C_{60}$, is demonstrated. The concept of orbital frustration explains the electronic structure of the elemental CKL and its direct band gap as a result of a metal-insulator transition. The unique combination of electronic and optoelectronic properties of the CKL would make it a superb semiconductor for applications. Possible routes to experimentally fabricate the CKL are also discussed.

Work in China was supported by the National Natural Science Foundation of China (Nos. 51176161, 51376005 and 11204262). Work at RPI was supported by US DOE under Grant No. DE-SC0002623. The supercomputer time was provided by NERSC under DOE Contract No. DE-AC02-05CH11231 and the CCI at RPI. MLC was supported by NSF Grant No. DMRIO-1006184 and the Theory Program at the Lawrence Berkeley National Laboratory through the Office of Basic Energy Science, US Department of Energy under Contract No. DE-ac02-05ch11231. We thank Susumu Saito and Yuki Sakai for useful discussions.

*Correspondence author:* [*]zhangs9@rpi.edu

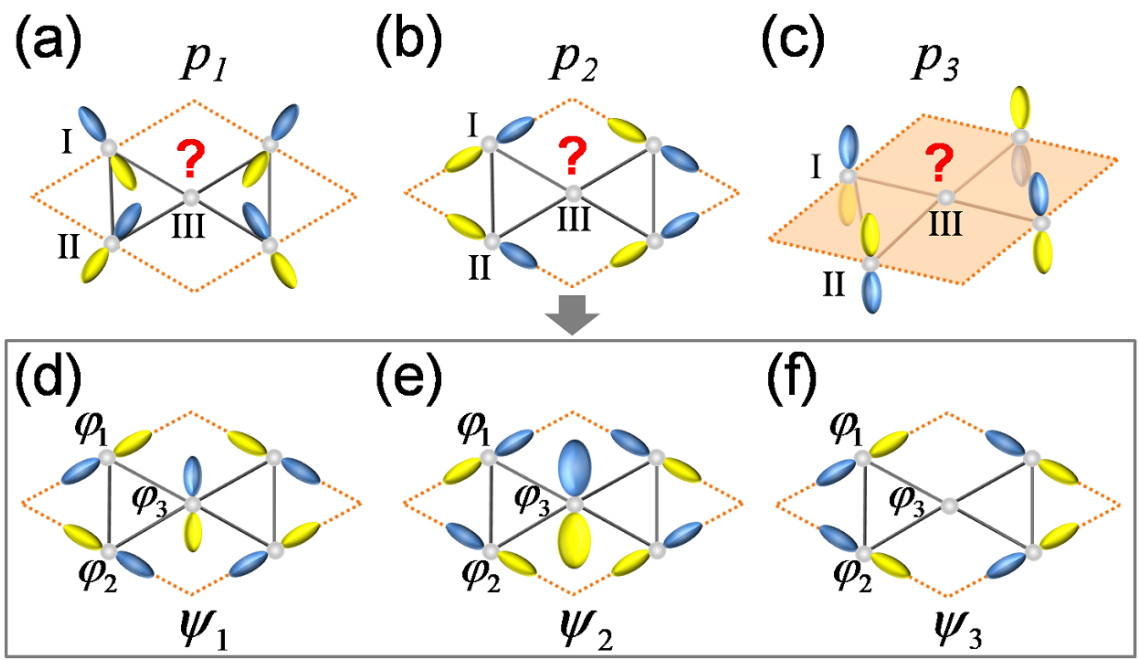

Figure 1 (color online). Orthogonal $p$ orbitals on a Kagome lattice. (**a-b**) In-plane $p_1$ and $p_2$ and (**c**) out of plane $p_3$ orbitals. (**d-f**) Three states derived from the $p_2$ orbital by applying symmetry operations, corresponding to the wavefunctions $\psi_1$, $\psi_2$ and $\psi_3$, respectively. I, II and III are three lattice sites of the primitive cells (dotted rhombic lines). Yellow and blue ellipsoids represent negative and positive lobes of $p$ orbitals, respectively.

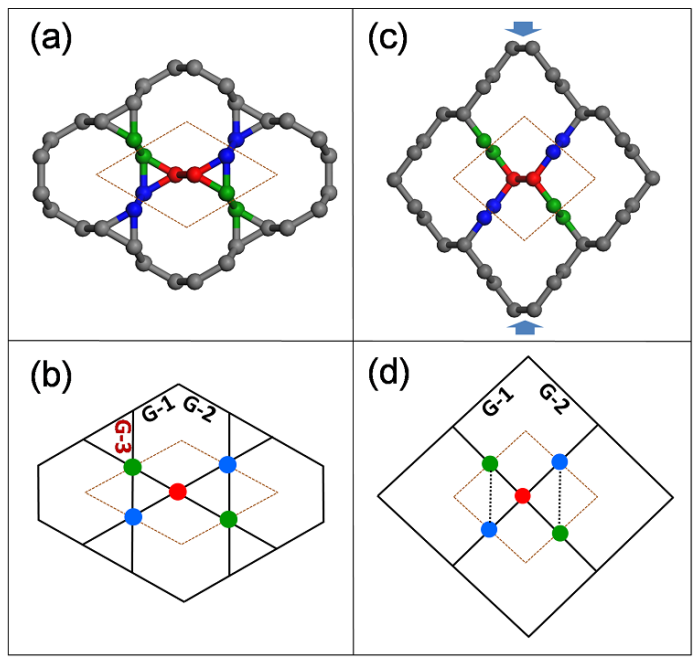

Figure 2 (color online). Crystal structures. (**a**) CKL. The unit cell consists of six C atoms in the form of two linked triangles. Each pair of (same color) atoms form a zigzag chain in the vertical direction. (**b**) A schematic Kagome lattice for the CKL, where G-1, G-2, and G-3 form three interlocked graphene lattices. Note that here each zigzag chain in the real structure is condensed to a lattice point. (**c-d**) same for the IGN with same legends. Notice the disappearance of G-3 in the IGN.

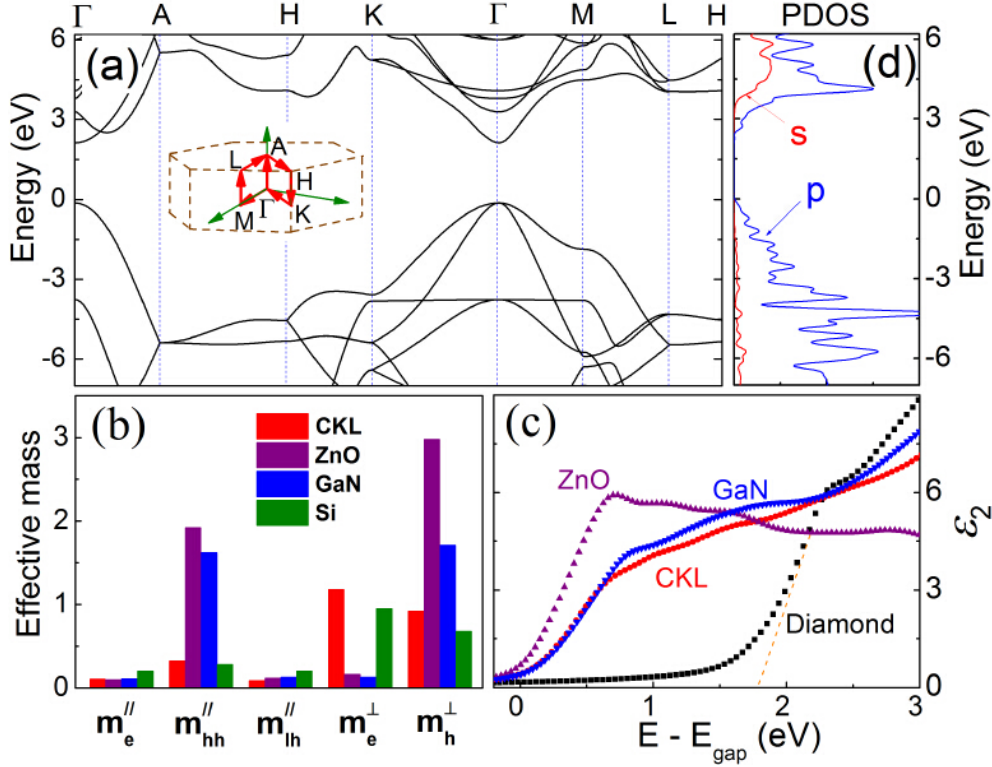

Figure 3 (color online). Electronic and optical properties. (**a**) Band structure of CKL. (**b**) Effective masses of CKL, ZnO, GaN, and Si calculated using DFT-PBE. For CKL, ZnO, and GaN, $m_e^{//}$ and $m_e^{\perp}$ are the in-plane and c-axis effective masses of electrons, $m_{hh}^{//}$ and $m_{lh}^{//}$ are the in-plane effective masses of heavy and light holes, whereas $m_h^{\perp}$ is the c-axis effective mass of heavy hole. For Si, $m_e^{//}$ and $m_e^{\perp}$ are the transverse and longitudinal masses of electrons, $m_{hh}^{//}$ and $m_{lh}^{//}$ are the effective masses of heavy and light holes along [001], whereas $m_h^{\perp}$ is the effective mass of heavy hole along [111] [40]. (**c**) Imaginary part of dielectric function $\varepsilon_2$ as a function of $E - E_{gap}$ for CKL, ZnO, GaN, and diamond, where $E_{gap}$ is the band gap. (**d**) Partial density of states (PDOS) of CKL, in which red and blue lines represent *s* and *p* states, respectively.

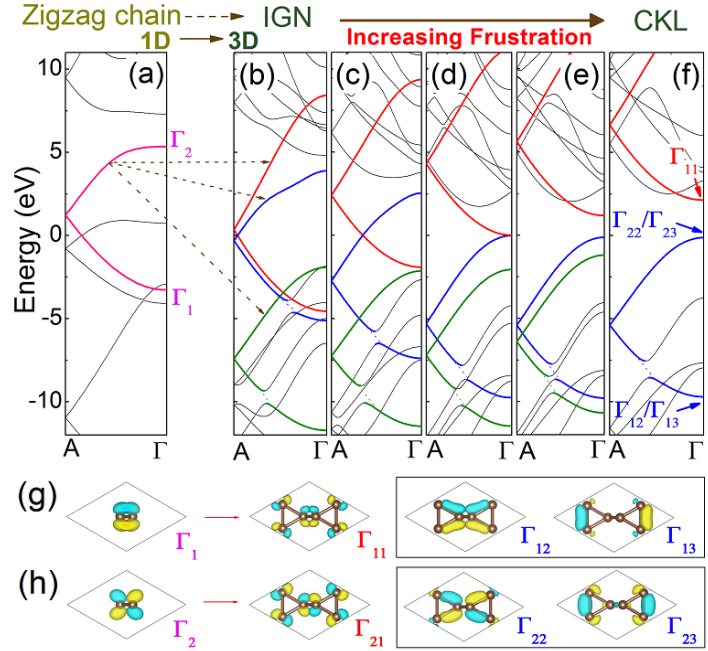

Figure 4 (color online). Band structure evolution with increased orbital frustration. (**a**) Zigzag carbon chain, (**b**) IGN, (**c-e**) intermediate structures, and (**f**) CKL. Colored bands in (**a**) split into three (red, blue and green) bands in (**b-f**). (**g**) Wavefunctions for $\Gamma_1, \Gamma_{11}$, and $\Gamma_{12}/\Gamma_{13}$ states, respectively. Note the correspondences: $\Gamma_{11} \rightarrow$ Fig. 1(**d**), $\Gamma_{12} \rightarrow$ Fig. 1(**e**), and $\Gamma_{13} \rightarrow$ Fig. 1(**f**). Also, having two atoms at each Kagome lattice site, the $\Gamma_{11}$ state here appears "fatter" than the one in Fig. 1(**d**). The banana shaped contours for $\Gamma_{12}/\Gamma_{13}$ are a result of in-phase addition of the schematic wavefunctions in Figs. 1(**e-f**). (**h**) Wavefunctions for $\Gamma_2$, $\Gamma_{21}$ (at $E = 13.9$ eV in (**f**)), and $\Gamma_{22}/\Gamma_{23}$.

# Supplemental Material

## 1. Derivation of orbital distribution in the Kagome lattice

We consider the case for orbital $p_2$ in Fig. 1(b). Let the orbitals on atoms I, II and III be $\varphi_1$, $\varphi_2$ and $\varphi_3$, respectively, as shown in Fig. 1(d). We apply the $B_{1u}$ projection operator $P_{B_{1u}}$ to one of the three orbitals, say $\varphi_3$. In particular, we apply each symmetry operation in turn, multiply the result by the character of the $B_{1u}$ representation, and then add the results up:

$$\psi_1 = P_{B_{1u}}\varphi_3 = \varphi_3 + \varphi_2 + \varphi_1 + \varphi_1 + \varphi_2 + \varphi_3 + \varphi_3 + \varphi_2 + \varphi_1 + \varphi_3 + \varphi_2 + \varphi_1 + \varphi_3 + \varphi_2 + \varphi_1$$
$$+ \varphi_1 + \varphi_2 + \varphi_3 + \varphi_3 + \varphi_2 + \varphi_1 + \varphi_3 + \varphi_2 + \varphi_1$$
$$= 8(\varphi_1 + \varphi_2 + \varphi_3).$$

After normalization, we obtain $\psi_1 = \dfrac{1}{\sqrt{3}}(\varphi_1 + \varphi_2 + \varphi_3)$, whose orbital configuration is shown in Fig. 1(d). The character table indicates that there should be two degenerate states corresponding to the $E_{1u}$ irreducible representation. We apply the $E_{1u}$ projection operator $P_{E_{1u}}$ to orbital $\varphi_3$ to obtain

$$\psi_2 = \dfrac{1}{\sqrt{6}}(2\varphi_3 - \varphi_1 - \varphi_2),$$

which corresponds to the orbital configuration in Fig. 1(e). In order to get $\psi_3$, we pick a symmetry operation in the $D_{6h}$ group (for example, $C_3$), which will convert $\psi_2$ into a different state and then make a linear combination of the new state with $\psi_2$, such that the result is orthogonal to $\psi_2$,

$$\psi_3 = 2C_3(\psi_2) + \psi_2 = \sqrt{\dfrac{3}{2}}(\varphi_2 - \varphi_1).$$

After normalization we obtain $\psi_3 = \dfrac{1}{\sqrt{2}}(\varphi_2 - \varphi_1)$, as shown in Fig. 1(f). The two degenerate states in Figs. 1(e) and 1(f) have either bonding or antibonding character.

## 2. Supplementary figures and tables

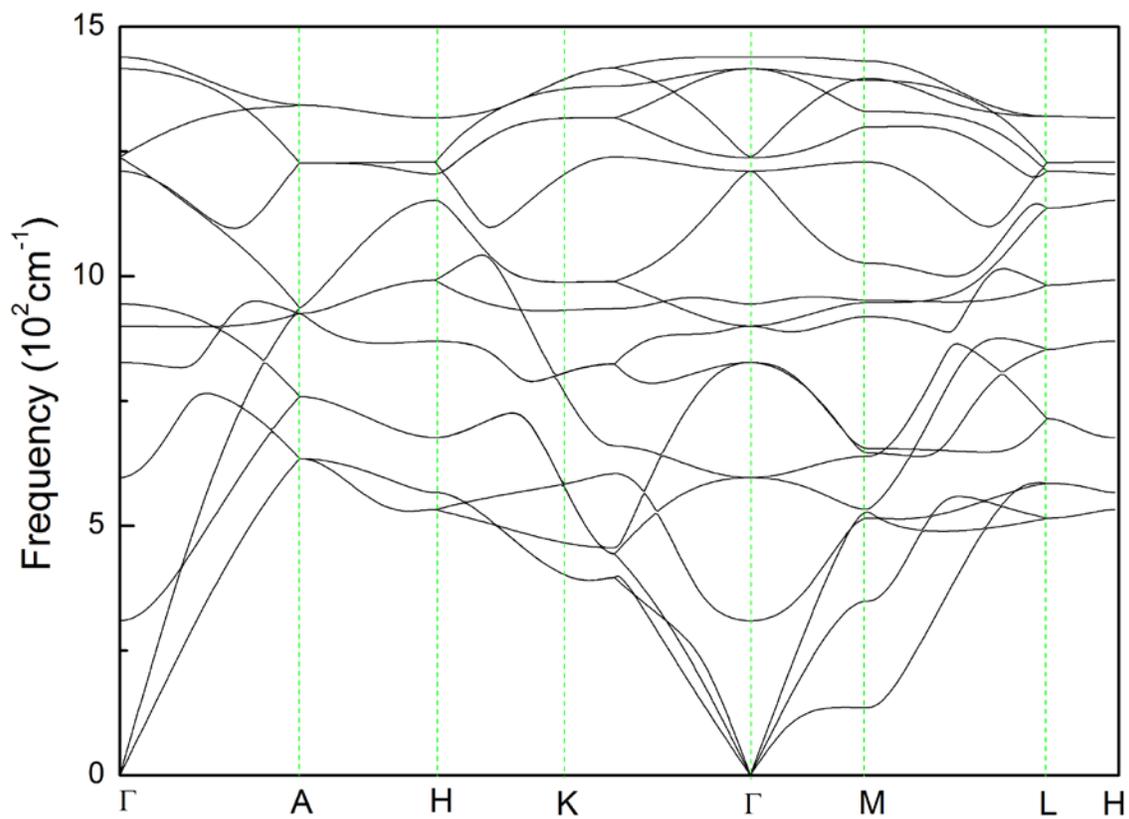

Figure 1 Phonon dispersion of CKL. The highest optical phonon frequency at zone center is 1440 cm$^{-1}$, which is lower than that of graphene (1560 cm$^{-1}$), but higher than that of diamond (1292 cm$^{-1}$). All frequencies are calculated using DFT-PBE.

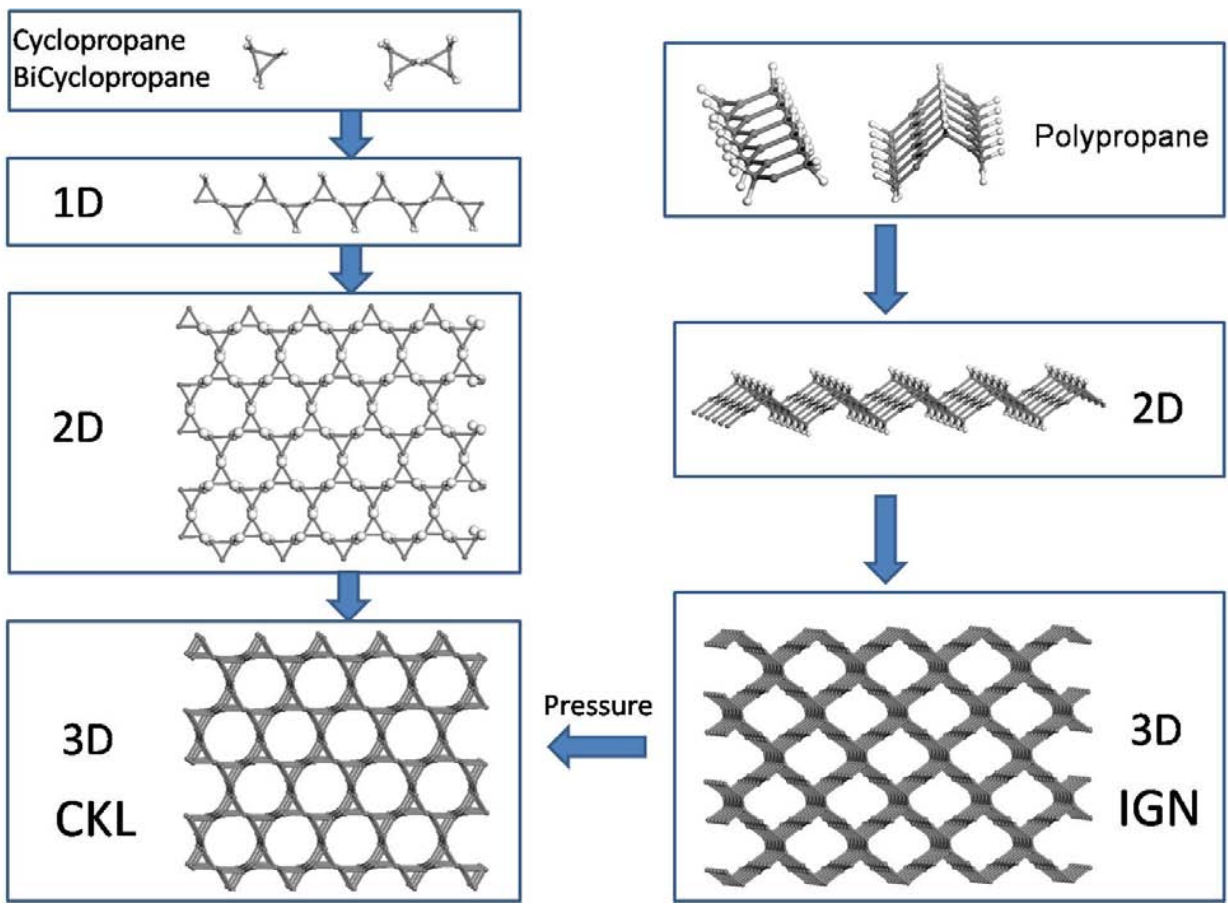

Figure 2 Possible routes for fabricating CKL. Left panel: the route from cyclopropane/bicyclopropane molecules to 1D chain and 2D sheet and then to 3D CKL. The 2D sheet is a single-layer planar structure after dehydrogenation. Right panel: the route from polypropane molecular chain to 2D zigzag sheet and then to 3D IGN. An uniaxial stress will convert the IGN to CKL.

|       | E | 2C$_6$ | 2C$_3$ | C$_2$ | 3C$_2'$ | 3C$_2''$ | i  | 2S$_3$ | 2S$_6$ | σ$_h$ | 3σ$_d$ | 3σ$_v$ |
|-------|---|--------|--------|-------|---------|----------|----|--------|--------|-------|--------|--------|
| A$_{1g}$ | 1 | 1  | 1  | 1  | 1  | 1  | 1  | 1  | 1  | 1  | 1  | 1  |
| A$_{2g}$ | 1 | 1  | 1  | 1  | -1 | -1 | 1  | 1  | 1  | 1  | -1 | -1 |
| B$_{1g}$ | 1 | -1 | 1  | -1 | 1  | -1 | 1  | -1 | 1  | -1 | 1  | -1 |
| B$_{2g}$ | 1 | -1 | 1  | -1 | -1 | 1  | 1  | -1 | 1  | -1 | -1 | 1  |
| E$_{1g}$ | 2 | 1  | -1 | -2 | 0  | 0  | 2  | 1  | -1 | -2 | 0  | 0  |
| E$_{2g}$ | 2 | -1 | -1 | 2  | 0  | 0  | 2  | -1 | -1 | 2  | 0  | 0  |
| A$_{1u}$ | 1 | 1  | 1  | 1  | 1  | 1  | -1 | -1 | -1 | -1 | -1 | -1 |
| A$_{2u}$ | 1 | 1  | 1  | 1  | -1 | -1 | -1 | -1 | -1 | -1 | 1  | 1  |
| B$_{1u}$ | 1 | -1 | 1  | -1 | 1  | -1 | -1 | 1  | -1 | 1  | -1 | 1  |
| B$_{2u}$ | 1 | -1 | 1  | -1 | -1 | 1  | -1 | 1  | -1 | 1  | 1  | -1 |
| E$_{1u}$ | 2 | 1  | -1 | -2 | 0  | 0  | -2 | -1 | 1  | 2  | 0  | 0  |
| E$_{2u}$ | 2 | -1 | -1 | 2  | 0  | 0  | -2 | 1  | 1  | -2 | 0  | 0  |

Table 1 Character table for the D$_{6h}$ point group.